\documentclass[preprint]{aastex}
\newcommand{\beq}{\begin{equation}}
\newcommand{\eeq}{\end{equation}}
\def\stacksymbols #1#2#3#4{\def\theguybelow{#2}
        \def\verticalposition{\lower#3pt}
        \def\spacingwithinsymbol{\baselineskip0pt\lineskip#4pt}
        \mathrel{\mathpalette\intermediary#1}}
\def\intermediary #1#2{\verticalposition\vbox{\spacingwithinsymbol
        \everycr={}\tabskip0pt
        \halign{$\mathsurround0pt#1\hfil##\hfil$\crcr#2\crcr
                \theguybelow\crcr}}}

\catcode`\@=11
\def\gsim{\ifmmode{\mathrel{\mathpalette\@versim>}}
    \else{$\mathrel{\mathpalette\@versim>}$}\fi}
\def\lsim{\ifmmode{\mathrel{\mathpalette\@versim<}}
    \else{$\mathrel{\mathpalette\@versim<}$}\fi}
\def\@versim#1#2{\lower 2.9truept \vbox{\baselineskip 0pt \lineskip 
    0.5truept \ialign{$\m@th#1\hfil##\hfil$\crcr#2\crcr\sim\crcr}}}
\catcode`\@=12

\def\Mbh{M_{\rm BH}}
\def\Mbht{\Mbh^{\rm T}}
\def\Nbh{N_{\rm BH}^>}
\def\Nq{N_{\rm Q}}
\def\Nqp{\Nq^>}
\def\Nqs{N_{\rm Q*}}
\def\Ng{N_{\rm g}}
\def\Ms{M_{\rm S}}

\def\Lq{L_{\rm Q}}
\def\Lqs{L_{\rm Q*}}
\def\Lqt{\Lq^{\rm T}}
\def\Etq{E^{\rm T}_{\rm Q}}

\def\Ledd{L_{\rm Edd}}
\def\Ls{L_{\rm S}}

\def\Lsni{L_{{\rm S*}i}}

\def\fq{f_{\rm Q}}
\def\fqN{f_{\rm Q,N}}
\def\fqM{f_{\rm Q,M}}

\def\tH{t_{\rm H}}

\def\epsq{\epsilon}

\def\alphaq{\alpha_{\rm Q}}

\def\Phiq{\Phi_{\rm Q}}

\def\Phis{\Phi_{\rm S}}

\def\Phisin{\Phi_{{\rm S*}i}}
\def\alphai{\alpha_i}
\def\Lsun{L_{\odot}}
\def\Msun{M_{\odot}}

\lefthead{Haiman, Ciotti, \& Ostriker}
\righthead{Local Black Holes and the Evolution of Quasars}

\begin{document}

\title{Reasoning From Fossils: Learning From the Local Black Hole
Population About the Evolution of Quasars}

\author{Zolt\'an Haiman\altaffilmark{1}, Luca Ciotti\altaffilmark{2,4,5}, 
\& Jeremiah P. Ostriker\altaffilmark{2,3}}

\affil{$^1$Department of Astronomy, Columbia University, 550 W120th St., New York, NY 10027, USA}

\affil{$^2$Princeton University Observatory, Princeton, NJ 08544, USA}

\affil{$^3$Institute of Astronomy, Cambridge University, Madingley Road, CB3 0HA, Cambridge, UK}

\altaffiltext{4}{On leave from Dipartimento di Astronomia,
Universit\`a di Bologna, via Ranzani 1, 40127 Bologna, Italy}

\altaffiltext{5}{Also at Scuola Normale Superiore, Piazza dei Cavalieri
7, 56126 Pisa, Italy}

\begin{abstract}

We discuss a simple model for the growth of supermassive black holes
(BHs) at the center of spheroidal stellar systems. In particular, we
assess the hypotheses that (1) star formation in spheroids and BH
fueling are proportional to one another, and (2) the BH accretion
luminosity stays near the Eddington limit during luminous quasar
phases.  With the aid of this simple model, we are able to interpret
many properties of the QSO luminosity function, including the puzzling
steep decline of the characteristic luminosity from redshift $z\approx
2$ to $z=0$: indeed the residual star formation in spheroidal systems
is today limited to a small number of bulges, characterized by stellar
velocity dispersions a factor of $2-3$ smaller those of the elliptical
galaxies hosting QSOs at $z \ga 2$. A simple consequence of our
hypotheses is that the redshift evolution of the QSO emissivity and of
the star formation history in spheroids should be roughly parallel. We
find this result to be broadly consistent with our knowledge of the
evolution of both the global star formation rate, and of the evolution
of the QSO emissivity, but we identify interesting discrepancies at
both low and high redshifts, to which we offer tentative solutions.
Finally, our hypotheses allow us to present a robust method to derive
the duty cycle of QSO activity, based on the observed QSO luminosity
function, and on the present--day relation between the masses of
supermassive BHs and those of their spheroidal host stellar systems.
The duty cycle is found to be substantially less than unity, with
characteristic values in the range $3-6\times 10^{-3}$, and we compute
that the average bolometric radiative efficiency is $\epsilon\approx
0.07$. Finally, we find that the growth in mass of individual black
holes at high redshift ($z\gsim 2$) can be dominated by mergers, and is
therefore not necessarily limited by accretion.

\end{abstract}


\section{Introduction}

The discovery of remarkable correlations between the masses of
supermassive BHs hosted at the centers of galaxies and the global
properties of the parent galaxies themselves (see, e.g., Magorrian et
al. 1998; Ferrarese \& Merritt 2000; Gebhardt et al. 2000; Graham et
al. 2001) begs for interpretation.  Several groups have noted the
natural link between the cosmological evolution of QSOs and the
formation history of galaxies (see, e.g. Monaco et al. 2000; Kauffmann
\& Haehnelt 2001; Granato et al. 2001; Ciotti \& van Albada 2001,
Cavaliere \& Vittorini 2002; Menci et al. 2003, and references
therein). The investigation of these interesting correlations looks
promising not only to yield a better understanding of how and when
galaxies formed, but also to obtain information about the QSO
population itself (Ciotti, Haiman, \& Ostriker 2001; Yu \& Tremaine
2002). For example, it may help us understand the well known but
puzzling fact that the characteristic QSO luminosity (obtained from
the QSO luminosity function, see, e.g., Pei 1995; Madau, Haardt \&
Rees 1999, Wyithe \& Loeb 2002) drops from $z\simeq 2.5$ to $z\simeq
0$ by a factor of $35\pm 15$.  On the face of it, this result is
surprising, since BHs can only grow, due to accretion or to mergers,
and more massive BHs are expected to be more luminous on average,
provided a sufficient amount of fuel is available.

Here we focus on a few specific points raised by the general remarks
above: (1) What drives the evolution of the steep decline with cosmic
time of the quasar luminosity density, and of the characteristic
quasar luminosity?  (2) What is the expected relation between the
cosmological evolution of the total emissivity in star--forming
galaxies and that of the total emissivity of the quasar population?
(3) How can one use scaling relations between the BH mass (hereafter
$\Mbh$) and the host galaxy properties to determine the QSO duty cycle
at redshift $z\simeq 0$?

The rest of this paper is organized as follows. In \S~2, we state our
hypotheses and we list the observational inputs required by our
approach.  In \S~3, we illustrate the technique adopted, and we
explore quantitatively its consequences by linking the star formation
history to the QSO evolution and applying it to explain the decrease
of QSO mean luminosity with decreasing redshift. Then, in \S~4, we
present robust estimates of the QSO duty cycle and derive the mean
accretion efficiency. Finally, in \S~5, we conclude by summarizing the
main results and the implications of this work.

Throughout this paper, we adopt the background cosmological model as
determined by the Wilkinson Microwave Anisotropy Probe ({\it WMAP};
Bennett et al. 2003) experiment.  This model has zero spatial
curvature, and is dominated by cold dark matter (CDM) and a
cosmological constant ($\Lambda$), with $\Omega_m=0.29$,
$\Omega_b=0.047$, and $\Omega_\Lambda=0.71$, a Hubble constant
$H_0=72~{\rm km~s^{-1}}$, an $rms$ mass fluctuation within a sphere of
radius $8 \; h^{-1}$ Mpc of $\sigma_8=0.9$, and power--law index
$n=0.99$ for the power spectrum of density fluctuations (Spergel et
al. 2003).  These values are consistent with their determinations by
most other methods (Bridle et al. 2003; Bahcall et al. 1999).

\section{Basic Assumptions and Model Ingredients}

A widely accepted consequence of the so--called Magorrian relation,
i.e., the (present--day) approximately linear relation between $\Mbh$
and $\Ms$, the host spheroid stellar mass, is that the bulk of BH
fueling in AGNs must be associated with star formation in the
spheroidal components of their host galaxies (Monaco et al. 2000; Page
et al. 2001; Granato et al. 2001; 2002; Cavaliere \& Vittorini 2002).
In this paper, we examine the simplest possible form of this
association, namely {\it the hypothesis that spheroid star formation
and BH fueling are -- at any time and in any system -- proportional to
one another with the proportionality constant independent of time and
system}.

Since most of the mass of BHs appears to have assembled within a
narrow redshift interval $\Delta z\approx 1$ around $z\approx 2$
(Boyle et al. 2000; Stoughton et al. 2002), in practice this
hypothesis needs to hold only during this redshift interval, in order
explain the local linear relation between BH and spheroid mass.  One
could argue that the energetic output from the forming central BH is
the driving physical process that at the end will establish the galaxy
mass (with the required proportionality). Alternatively, stars would
form first, and then the BH is formed from reprocessed gas. In this
case, a source of fuel for the BH growth with the required
proportionality (namely mass losses from the newly formed stars), is
available in a natural way.  One can imagine that both of the above
scenarios lead to a linear BH vs. spheroid mass relation at $z=0$, but
the strict proportionality in mass accretion rates into BHs and
spheroids may not hold at all redshifts.  Nevertheless, it is
interesting to ask whether the simple hypothesis above is consistent
with other observational data at both lower and higher redshifts,
where {\it some} mass is still being added to both BHs ad spheroids,
since this test can reveal information about the physical process of
the BH and spheroid mass assembly.

Since BH fueling should inevitably lead to some form of QSO activity,
we make {\it a second simple hypothesis, namely that the BH accretion
luminosity always stays near the Eddington limit when the QSO is in
the luminous, or ``on'' phase''}.  This is apparently different from
other proposals in the literature that are variants of the ``feast or
famine'' model (Small \& Blandford 1992), and which posit that QSO
activity declines towards redshift $z=0$ owing, at least in part, to a
significant decrease in the fueling rate (Cavaliere et al. 2000;
Haiman \& Menou 2000; Kauffmann \& Haehnelt 2000). In fact overall
``activity'', i.e. the luminosity density evolution of QSOs, is the
product of their characteristic number density $\Nqs (z)$ and their
characteristic luminosity $\Lqs (z)$. 
The product $\Nqs \Lqs$ may decline due to a decline in fueling that
shuts off AGN activity and primarily leads to a decline in $\Nqs$. But
it is a separate question to ask what causes the surprising, but well
observed decline with increasing time (Pei 1995; Boyle et al. 2000;
Stoughton et al. 2002) in the characteristic luminosity $\Lqs (z)$.

Coupled with the Magorrian relation, the above two hypotheses allow us
to make several simple predictions, that will be described in detail
in the following sections. Before we present our results, we list in
detail the observational inputs required by our approach.

The {\it first observational input} of our analysis is the {\it
present--day} luminosity function (hereafter LF) of spheroids, where
the number of spheroids per unit volume with luminosities in the
interval $(\Ls, \Ls+d\Ls)$ is defined to be given by $\Phis
(\Ls)d\Ls$.  A composite LF was presented recently by Salucci et
al. (1999), who considered the LF of four different types (E, S0,
Sa/Sab, and Sbc/Scd) of galaxies separately, and inferred the total
spheroid LF by assuming that on average, the spheroid components
contribute 90\%, 65\%, 40\%, and 10\% of the light of the above
galaxies, respectively.  The composite spheroid LF is therefore
represented by the sum of four different ``Schechter-law''
distributions
\begin{equation}
\Phis(\Ls)=\sum_{i=1}^4 \frac{\Phisin}{\Lsni}\times 
\left({\Ls\over \Lsni}\right)^{-\alphai}\exp \left(-\frac{\Ls}{\Lsni}\right),
\label{eq:dphisphdl}
\end{equation}
where $\log\ (\Phisin /{\rm Gpc^{-3}})=5.89, 5.95, 6.03, 6.45$, $\log\
(\Lsni/{\rm L_\odot}) = 10.18, 10.02, 10.10, 9.90$, and $\alphai=0.95,
0.95, 1.0, 1.3$, for E and the bulges of S0, Sa/Sab, Sbc/Scd galaxies,
respectively.  Benson, Frenk, \& Sharples (2002) have recently derived
the spheroid luminosity function for a small sample of 90 bright field
galaxies by decomposing the bulge and disk components, while Bernardi
et al. (2002) and Sheth (2003) have computed (see also Yu \& Tremaine
2002) the velocity function of early--type galaxies in the SDSS.
While neither of these can serve as a substitute for the full spheroid
luminosity function to replace equation~(\ref{eq:dphisphdl}) above,
this should undoubtedly be possible in the near future by decomposing
a large sample of fainter late--type SDSS galaxies into their bulge
and disk components.

The {\it second ingredient} is the quasar LF and its evolution with
redshift,
\begin{equation}
\Phiq(\Lq,z)={\Phi_{\rm Q*}/L_{\rm Q*}(z) \over [\Lq/L_{\rm
Q*}(z)]^{\beta_l}+[\Lq/L_{\rm Q*}(z)]^{\beta_h}}:
\label{eq:phiq} 
\end{equation}
the optical data in the rest--frame B band can be well fitted by pure
luminosity evolution, with the characteristic luminosity $L_{Q*}$
evolving with redshift as
\begin{equation}
L_{Q*}(z)=L_{Q*}(0)(1+z)^{\alphaq-1}~{e^{\zeta z}(1+e^{\xi z_*})\over 
e^{\xi z}
+ e^{\xi z_*}}. 
\label{eq:lqz}
\end{equation}
We adopt the fitting parameters given by Madau, Haardt \& Rees (1999),
$\beta_l=1.64$, $\beta_h=3.52$, $z_*=1.9$, $\zeta=2.58$, $\xi=3.16$
and $\alphaq=0.5$.  The characteristic space density and luminosity
are provided by Pei (1995) in a standard CDM cosmology with
$H_0=50~{\rm km~s^{-1}}$ as $\log\ (\Phi_{Q*}/{\rm Gpc^{-3}})=2.95$
and $\log\ [L_{Q*}(0)/\Lsun]=13.03$: we adopt these values with
appropriate redshift--dependent re--scalings to our $\Lambda$CDM
cosmology.

Finally, the {\it third ingredient} is the Faber-Jackson relation (1976) 
\begin{equation}
\frac{\Ls}{10^{11}\Lsun} \simeq 0.62 
\left (\frac{\sigma}{\rm 300 km~s^{-1}}\right)^{4.2},
\label{eq:FJ}
\end{equation}
in the relatively more recent version of Davies et al. (1983),
coupled with the $\Mbh-\sigma$ relation (Ferrarese \& Merritt 2000,
Gebhardt et al. 2000, Tremaine and Yu 2002),
\begin{equation}
\frac{\Mbh}{\rm 10^9 M_\odot} \simeq 
\left(\frac{\sigma}{\rm 300 km~s^{-1}}\right)^4.
\label{eq:Magoo}
\end{equation}
Equation~(\ref{eq:FJ}) is only approximately true, and the slope turns
shallower for galaxies with velocity dispersions below\footnote{For
example, a fit to galaxies in Virgo cluster by Dressler et al. (1987)
gives a mean exponent of $\simeq 3.5$, while Bernardi et al. (2003)
found instead a value of $\simeq 4$ for the exponent.}  $\sigma \la
170\ {\rm km\, s^{-1}}$. Likewise, the exponent in
equation~(\ref{eq:Magoo}) is currently under debate (Ferrarese \&
Merritt 2000; Gebhardt et al. 2000); here we refrain from a critical
assessment of the different values found in the literature, and accept
the slope $\sim 4$ as approximately the true value for both relations.
Thus, to a good (but not necessarily perfect) accuracy, both the $\Mbh
-\sigma$ and the Faber-Jackson relations indicate a proportionality to
the fourth power of the central velocity dispersion, implying the
following linear relation:
\begin{equation}
\frac{\Mbh}{\Msun} \simeq 0.016\frac{\Ls}{\Lsun}.
\label{eq:MbhLs}
\end{equation}
When expressing equation~(\ref{eq:MbhLs}) above in term of galaxy mass
instead of luminosity, it is found that the implied median BH mass
fraction to stellar mass is 0.13\% of the mass of the bulge (Kormendy
\& Gebhardt 2001).

\section{Linking the Star Formation History and the Evolution of Quasars}

As emphasized above, it is widely believed that the bulk of BH fueling
in AGNs must be associated with star formation in the spheroidal
components of their host galaxies.  In this section, we examine the
hypothesis stated in \S~2 above, namely that {\it spheroid star
formation and BH fueling are -- at any time and in any system --
proportional to one another with the proportionality constant
independent of time and system}.  Under the assumption that quasars
radiate a fixed fraction $\epsilon$ of their accreted mass, an obvious
consequence is that the redshift evolution of the QSO emissivity and
of the star formation history in spheroids should be roughly parallel
to each another. As we shall see, we find this result to be broadly
consistent with our knowledge of the evolution of both the global star
formation rate, and of the evolution of the QSO emissivity, but we
identify interesting discrepancies at both low and high redshifts, to
which we offer tentative solutions.

\begin{figure}
\begin{center}
\includegraphics[width=0.7\textwidth]{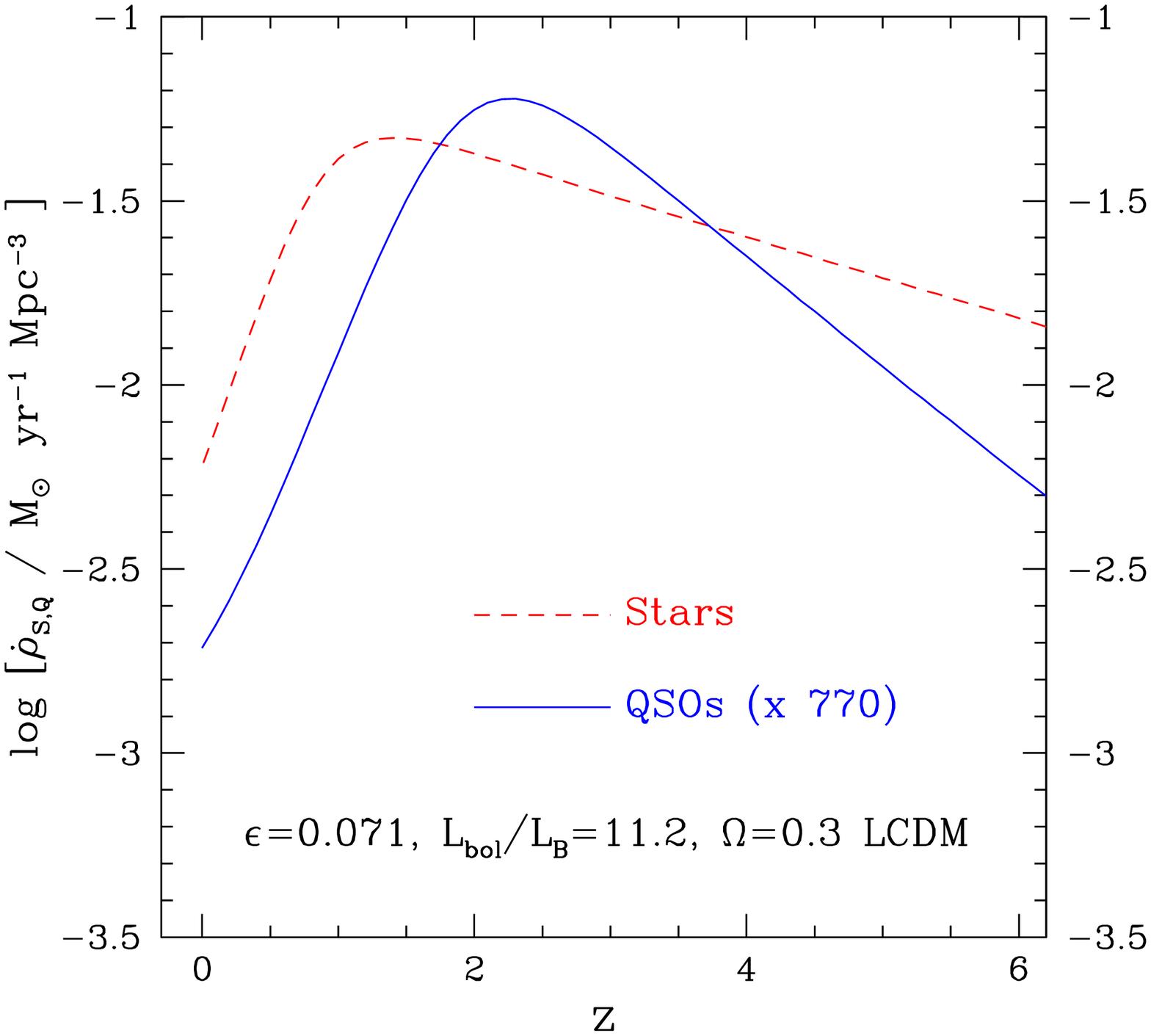}
\end{center}
\figcaption[fig1.eps]{The redshift evolution of the total star
formation rate density (SFRD; dashed curve) and the total black hole
accretion rate density (BARD; solid curve).  The SFRD was adopted from Madau \&
Pozzetti (2000); while the BARD is obtained from the optical quasar
luminosity function, assuming a bolometric correction of $A_{\rm
bol}=11.2$ and a constant radiative efficiency of $\epsilon=0.071$
(independent of redshift and quasar luminosity).  The BARD is
displaced upward by a constant factor of $1/\epsilon=770$ for clarity
of presentation.
\label{fig:madau}}
\end{figure}

The evolution of the total UV luminosity density in stars at 1500\AA\
(galaxy rest frame) with redshift is given (in a standard CDM
cosmology with $H_0=50~{\rm km~s^{-1}}$; Madau \& Pozzetti 2000) by
\beq 
\dot\rho_{\rm S,UV}=7\times10^{26} 
\frac{\exp(3.5 z)}{\exp(3.75 z)+20} \,\,\,{\rm erg\, s^{-1} Hz^{-1} Mpc^{-3}}.
\label{eq:rhosuv}
\eeq
This is related to the total star formation rate density as 
\beq
\dot\rho_{\rm S}=\frac{\dot\rho_{\rm S, UV}}{8\times 10^{27}} \,\,\,{\rm M_\odot\, yr^{-1} Mpc^{-3}}
\label{eq:sfr}
\eeq for a Salpeter IMF (Madau et al. 1998).  Figure~\ref{fig:madau}
shows (dashed curve) this star formation rate density (SFRD), with and
appropriate redshift--dependent re--scaling to our adopted
$\Lambda$CDM cosmology.  This SFRD is close to that derived more
directly in the recent work by Porciani \& Madau (2001).

The evolution of the total rest--frame B--band luminosity density in
quasars can be obtained from equations~(\ref{eq:phiq}) and
(\ref{eq:lqz}) as 
\beq 
j_{\rm Q, B}=\int_0^\infty dL\, L \Phiq(L,t).
\label{eq:rhoqb}
\eeq Under the assumption that quasars radiate a fixed fraction
$\epsilon$ of their accreted mass (see discussion below), this is
related to the total BH accretion rate density ($\dot\rho_{\rm Q, B}$,
hereafter BARD) as 
\beq j_{\rm Q}=\frac{A_{\rm bol}\epsilon c^2}{1-\epsilon} 
\left(\frac{\dot\rho_{\rm Q, B}}{\rm M_\odot\, yr^{-1} Mpc^{-3}}\right),
\label{eq:bhfr}
\eeq where $A_{\rm bol}=11.2$ is the bolometric correction, and
$\epsilon=0.071$ is the radiative efficiency, derived in the next
section.  Figure~\ref{fig:madau} shows (solid curve; displaced upward
by a constant factor of 770 for clarity) the BH mass accretion rate
density, re--scaled to our adopted $\Lambda$CDM cosmology.

As is well known, both the SFRD and the BARD exhibit a steep rise from
$z=0$ to $z=1-2$, a peak at $z\sim 1-2$, and a decline towards still
higher redshifts.  This is broadly consistent with their expected
parallel evolution under our simple set of assumptions.  Both the SFRD
and the BARD still have significant observational uncertainties. While
the steep decline at low redshift is relatively secure, the current
SFRD and the BARD determinations could both turn out to be
underestimates at high redshifts, due to yet--undetected populations
of galaxies or AGNs (e.g. due to dust obscuration).  A critical review
of the uncertainties is beyond the scope of this paper; we here simply
take the current determinations at face value, and examine
discrepancies at both low and high redshifts from our simple model.

\subsection{Low Redshifts ($z\la 2$)}

\subsubsection{Why Does the Characteristic QSO Luminosity Evolve?}

We start by demonstrating that the bulk of BH formation, and
consequently the bulk of QSO activity, must have occurred in galactic
systems dominated by massive, luminous bulges. In fact, the spheroid
light distribution is known to approximately satisfy a Schechter--like
distribution (see, e.g., eq.~\ref{eq:dphisphdl}) 
\beq
f(\Ls)\propto\left(\frac{\Ls}{L_{\rm S*}}\right)^{-\alpha}
\exp\left(-\frac{\Ls}{L_{\rm S*}}\right)
\label{eq:dPhiSph}
\eeq 
with $\alpha\approx 1.2\pm 0.1$ (Salucci et al. 1999; Benson et
al. 2002; Bernardi et al. 2003). Then, from $\Mbh\propto\Ls$, it
follows that one half of the mass in BHs is in systems with luminosity
$\Ls/L_{\rm S*}\geq \ell_{1/2}$, where $\ell_{1/2}$ is defined by 
\beq
\int_0^{\ell_{1/2}} \ell^{-\alpha+1}\exp(-\ell) d\ell = \frac{\Gamma
(2-\alpha)}{2}, 
\eeq 

and for $\alpha=1.2$, this yields $\ell_{1/2}\simeq 0.5$. In a more
detailed computation, using the composite spheroid luminosity function
given in equation~(\ref{eq:dphisphdl}), we find that the luminosity
above which half of the integrated light is emitted corresponds to a
spheroid with luminosity $M_B\approx -20.5$, only a factor of $\sim
2.5$ fainter than the luminosity of the well--known giant elliptical
M87 ($M_B=-21.42$). Thus {\it the bulk of the mass density of BHs
reside in giant ellipticals, and, if the current situation is not
anomalous, the bulk of the growth of SMBHs must also have occurred
there (or in progenitor systems)}.  We also know that the bulk of star
formation in spheroidal systems took place as early as redshift $z
>2$, as indicated, for example, by the mean stellar ages in
ellipticals (Hogg et al. 2002; Bernardi et al. 2003), and of bulge
populations (e.g. Proctor et al. 2000; Ellis et al. 2001), and the
Butcher--Oemler or Gunn--Dressler effects (Margoniner et al. 2001).

At present, the disks of spiral galaxies dominate the global star
formation rate (Fukugita et al. 1998; Benson et al. 2002; Hogg et
al. 2002), and the mean age of stars in spiral systems is perhaps a
factor of two younger than that in spheroidal systems.  It follows,
given our hypothesis, that BH growth in the local universe is
dominated by relatively small bulges that live in galaxies denominated
as spirals. Fortunately, the hypothesized relation between nuclear
activity and star formation can be directly tested at low redshift.
For example, Percival et al. (2001) obtained morphological information
for the host galaxies of nine bright, ($M_V < - 25.5$) QSOs,
classifying six of them as ``disks'', and the remaining three as
``spheroids''.  The bulk of the local population of identified QSOs
live in disk dominated systems. The sample studied by Percival et
al. was approximately three magnitudes brighter than the
characteristic luminosity of the local population of QSOs (Pei 1995),
and almost all low luminosity AGNs are known to reside in spiral
systems.  Our conclusion would therefore likely be strengthened by a
survey going to magnitudes fainter than studied in the Percival et
al. work, closer to $\Lqs$.

It is natural to ask what the consequences would be of the hypothesis
that BH fueling, when it does happen, stays near the Eddington
limit. This assumption is not unrealistic: in fact, for a handful of
nearby AGNs, the BH masses can be directly estimated by reverberation
mapping (or by a cruder ``photoionization method''; see Wandel et
al. 1999), and for these sources, the Eddington ratio can be directly
inferred.  From the 19 nearby AGNs listed in Table 3 and in Figure 5
of Wandel, Peterson, \& Malkan (1999), one derives $L=(0.01-0.3)\Ledd$
for the Seyfert 1 objects, while the two QSOs have $L/\Ledd = 0.2$ and
0.3. The bolometric correction for very hard and IR/submillimeter
radiation, using the mean quasar spectrum as given by Elvis et
al. (1994), is about a factor of $\sim 3$, which would bring the
luminosities of the two QSOs in the Wandel et al. sample close to
Eddington limit.

The assumption of always maintaining the Eddington luminosity, if
applied to {\it an individual} BH, predicts an {\it increasing} QSO
luminosity $\Lq$ (in the ``on'' state), due to the trivial fact that
for every BH the mass is monotonically increasing.  However, this is
not necessarily in conflict with observations that describe the
evolution of the characteristic luminosity for a population of
quasars. Clearly, if all galaxies remained equally active, then the
mean luminosity inferred from the observed QSO luminosity function
would increase, but if the typical member of the population changes
with time the naive expectations may be incorrect.

Let us simply assume as an empirically verified fact that at the
present day, star formation is active primarily in disk dominated
systems. A decrease of a factor 20--50 in the characteristic quasar
luminosity $\Lqs$ would then be naturally obtained by combining the
Faber--Jackson ($L\propto \sigma^4$) and Magorrian ($\Mbh \propto
\sigma^4$) relations with a reduction in the characteristic central
velocity dispersion ($\sigma$) in the hosts of QSOs by a factor in the
range 2.1--2.7.  A decrease of this amount is quite natural, when one
considers the mean (luminosity weighted) central velocity dispersion
associated with the ellipticals at redshift $z=2$ ($\sigma\approx
400~{\rm km~s^{-1}}$), and that associated with spiral bulges at
redshift $z=0$ ($\sigma\approx 200~{\rm km~s^{-1}}$), as derived by
the Faber-Jackson relation. This argument thus provides a
straightforward interpretation of why the typical QSO luminosity
decreases from $z=2$ to $z=0$.  Furthermore, there is explicit
observational evidence (Thomas et al. 2002) that large ellipticals are
older than small bulges of spirals, supporting the decrease in
$\sigma$ towards $z=0$ as the reason behind the decrease in the
characteristic quasar luminosity.

A slightly less steep drop in the central velocity dispersion could be
acceptable, by simultaneously allowing for quasar luminosities to
decrease to somewhat sub-Eddington values towards low redshifts.  The
latter scenario is consistent with the Eddington ratios of the two
quasars in the Wandel et al. sample. Applying the reverberation
mapping technique to an extended quasar sample (e.g. selected from the
Sloan Digital Sky Survey, SDSS) would distinguish directly among these
two options.

\subsubsection{Why Does the Total Quasar Emissivity Evolve?}

We next consider whether the observed steep evolution of the total
quasar emissivity (or BARD) is consistent with the star formation rate
density (SFRD).  Figure~\ref{fig:madau} shows the BARD and
SFRD. However, the SFRD includes contributions from both disk stars
and spheroid stars: here we discuss corrections to this diagram to
obtain the SFRD in spheroids alone.

Corrections for disk star formation are likely to become large at low
redshifts.  The fraction of the total stellar luminosity density at
$z=0$ contributed by stars in disks vs. stars in spheroids (the latter
including both the bulges of spirals and ellipticals) has been
estimated by several authors.  Fukugita et al. (1998) and Hogg et
al. (2002) find that spheroids contribute $\sim 40\%$ of the
luminosity density (but Benson et al. 2002 find spheroids to
contribute significantly less than this fraction).  Furthermore, it is
well--known that the stellar populations in present--day spheroids are
old, and most models place their formation epochs at $z>2$.
Nevertheless, we are interested in the amount of ongoing starformation
in these spheroids at $z=0$.  The lower limits on the ages of the
spheroid stellar populations come from various methods; one of these
is the colors.  $B-V$ and $V-I$ magnitudes can be determined to an
accuracy of $\sim 0.1$ mag (e.g. Ellis et al. 2001), and these colors
typically change by $\sim 1$ mag in a Gyr of evolution (Leitherer et
al. 1999).  It follows that $<10\%$ of the present--day luminosity in
these systems can arise from young stars. In turn, this implies that
less than $\sim 5\%$ of the total SFRD at $z=0$ is occuring in
spheroid systems (i.e. in the bulges of spirals).  On the other hand,
at redshifts of $z=2-4$, the Lyman break galaxies are believed to be
large bulge systems in the process of formation, and the fraction of
the starformation seen at these redshifts associated with bulges
essentially unity.

The Butcher-Oemler (also known as Dressler-Gunn) effect can also be
used to derive the contribution of spheroids to the total SFR as a
function of redshift in galaxy clusters. The BO effect then gives the
rate of expected decline in the emission from SMBHs, since these tend
to be in high mass ellipticals which are well represented in the BO
clusters.  Observational work on the BO effect has shown that the
number fraction of blue galaxies in clusters, $f_b$, increases with
$z$ in a linear relation (see, e.g., Newberry et al. 1988, Andreon \&
Ettori 1999, Metveier et al 2000).  For example, in one of the most
detailed works, based on the analysis of 295 POSS-II clusters with
redshift $0<z<0.4$ (Margoniner et al. 2001), the authors found
\begin{equation}
f_b\simeq 1.3^{\pm 0.5}z +c,
\end{equation}
where $c$ is a small additive constant, of the order of $0.02\pm 0.01$
There are important caveats in using the BO effect for the spheroid
correction. First, there are large variations in $f_b$ from cluster to
cluster (e.g. the $z=0.83$ cluster studied by Van Dokkum et al. (2000)
has an estimated $f_b = 0.22 \pm 0.09$).  It is also not clear at the
present time where the star formation responsible for this blue light
is occurring.  While this blue light may represent ongoing star
formation in spheroids, Abraham et al. (1996) argue that the blue
light arises in the disks that flare up as spirals fall into the
cluster potential.  However, interpreting the blue fraction $f_b$ as
the fraction of ellipticals undergoing starformation, the BO effect
would support the general conclusion that spheroids contribute only a
few percent of the total starformation rate in the present--day
universe, while this fraction rises steeply towards higher redshifts.

There are other promising methods to estimate the spheroid
contribution to the total SFRD. For example, one could measure local
starburst activity in the dense obscured centers of galaxies in the
infrared bands, and identify this with the local star formation rate
in bulges.  It should also be possible to measure an accurate age
distribution of stellar populations in the bulges of spirals, as well
as in elliptical galaxies, in large samples of SDSS galaxies, and
hence to directly infer the time dependence of the star formation rate
in spheroids as a function of redshift.  An accurate measurement of
the age distribution has already been achieved for a red subsample of
SDSS galaxies (Jimenez et al. 2003).

\begin{figure}
\begin{center}
\includegraphics[width=0.7\textwidth]{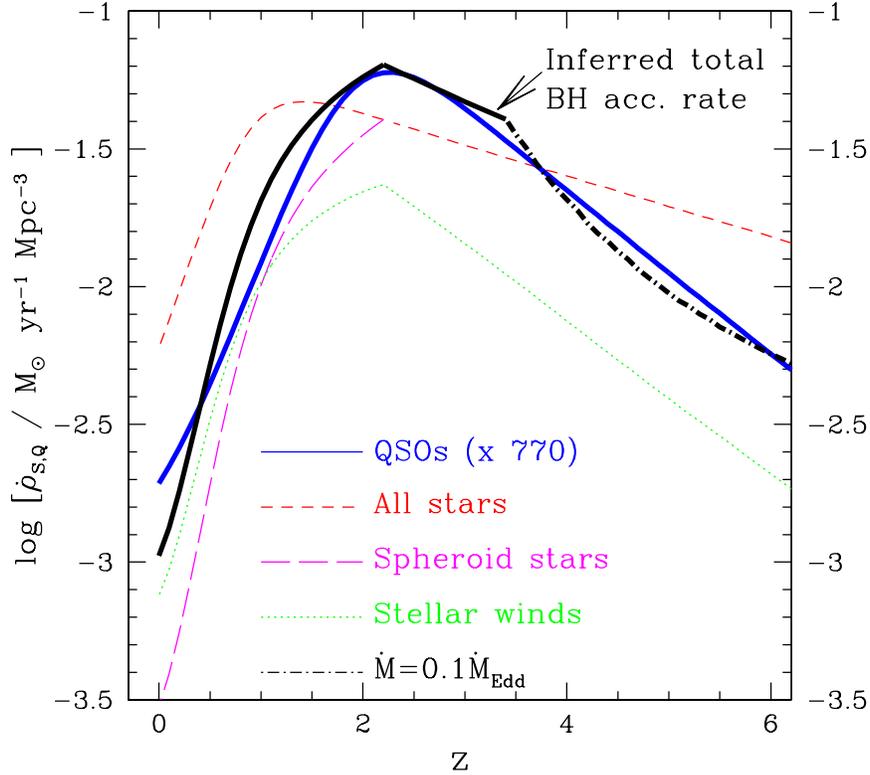}
\end{center}
\figcaption[fig2.eps]{Modifications to the SFRD and BARD in
Figure~\ref{fig:madau} that would bring the two quantities to parallel
each other; as required by our simple set of assumptions.  The
modifications include (1) a correction for the spheroid vs. total SFRD
(long--dashed curve); (2) the possibility of fueling BHs via stellar
winds (dotted curve); and (3) a suppression of the BARD at high
redshifts due to an extra source of opacity (dot--dashed curve). The
thick solid curve (turning into the dot--dashed curve at high
redshift) shows the BARD inferred from the SFRD after these
corrections are taken into account; it tracks the BARD inferred from
the optical quasar luminosity function (thin solid curve).
\label{fig:corr}}
\end{figure}

In summary, we here estimate the rough fraction of the observed
starformation rate that is associated with spheroids at each redshift
by multiplying the total SFRD shown in Figure~\ref{fig:madau} by a
factor $f_{\rm sph}=0.05+0.95 (z/2.2)^2$ at $z<2.2$.  This ensures a
smooth transition in the total SFRD being dominated by ellipticals at
high redshift to it being dominated by disks of spirals at $z=0$, with
only residual starformation in the bulges of spirals, consistent with
the arguments above.  In Figure~\ref{fig:corr}, we show (long--dashed
curve) the corrected SFRD.  As the figure demonstrates, including this
correction improves the fit, in the sense of making the SFRD resemble
the BARD more closely.  However, intriguingly, it does appear that the
decline in the spheroid formation rate from $z=2$ to $z=0$ is too
large, by a factor of $\sim$three, when compared to the decline in the
BARD. If this discrepancy holds up in future data, it would imply that
the nuclear black holes can be fueled long after the star formation in
the bulge has ceased.  Since the bulge star formation has likely used
up all the gas initially present in the bulge, the fuel would have to
arrive from elsewhere.

A simple assertion is that old stars that had formed in the bulge keep
returning a fraction of their mass in winds. These stellar winds may
provide dense, shocked material that can dissipate and serve as a fuel
for the central BH.  The mass--loss rate in winds in a starburst
evolves as $\propto t^{-1.3}$ (where $t$ is the time elapsed from the
burst; see, e.g., Ciotti et al. 1991; Leitherer et al. 1999).  We here
use the wind mass--loss rate $\dot M_{\rm wind}=1.5\times10^{-11} L_B
t_{15}^{-1.36}\,{\rm M_\odot yr^{-1}}$ between the ages of 0.5 and 15
Gyr for a one solar mass model starburst galaxy, where $t_{15}$ is the
time elapsed from the burst in units of 15 Gyr, and $L_B\approx 0.03$
is the B--band luminosity of the model galaxy (in units of $L_\odot$)
at 15 Gyr (Bruzual \& Charlot 2000). Under these assumptions, $\sim
80\%$ of the stellar mass is eventually returned to the ambient
medium.

For our purposes, we regard mass lost in winds as new material
available to fuel the central BH.  It is clear that the total mass
return rate from a passively evolving stellar population cannot
accrete onto the central BH (otherwise BH masses will be two or three
order of magnitude larger than those observed); nor can it all turn
into stars (star formation in present day ellipticals is not detected
at the level that would be implied). In order to solve these problems
Ciotti \& Ostriker (2001) performed numerical simulations of radiative
feedback modulated accretion flows onto a SMBH at the center of a
"cooling flow" galaxy.  They showed that only a {\it few percent (or
less)} of the available gas lost in winds effectively accretes onto
the central BH, while the accretion luminosity during short episodes
of bursts stays near the Eddington value (a similar conclusion would
follow in the case of mechanical feedback; e.g. Tabor \& Binney 1993;
Binney \& Tabor 1995; Binney 1999).

Here we assume that a fraction $1.3\times 10^{-3}$ of the mass in
winds accretes onto the central BH; i.e. the same fraction we had
assumed for the ``original'' infalling gas earlier in this paper.
This fits in well explicitly with the fractions inferred in Ciotti \&
Ostriker (2001).  In Figure~\ref{fig:corr}, we show (dotted curve) the
total mass loss rate generated in winds from spheroid stars as a
function of redshift.  The thick solid curve (turning into the
dot--dashed curve at high redshift, see discussion below) shows the
total BARD inferred from the SFRD after mass loss from winds are
added. We conclude that, if a significant fraction of the wind
material ends up fueling the central BH, this brings the BARD and SFRD
into quite reasonable agreement (to within a factor of two at all
redshifts).

We emphasize that our treatment in this subsection is
phenomenological, and complementary to theoretical semi--analytical
models (Haiman \& Menou 2000) of the cosmological evolution of the QSO
luminosity, or models based on Monte Carlo realizations of dark matter
``merger trees'' (see, e.g., Kauffmann \& Haehnelt 2001). These models
have found that to reproduce the observed decline in the QSO
luminosity density, the ``efficiency factor'' for the fraction of gas
accreted by the BH in a merger must decline towards $z=0$.  Our
proposal here is radically different: instead of ``starving'' the BHs
in each galaxy, the QSO luminosity density drops due to the
empirically--inferred drop in the formation rate of spheroids, and
their bias towards smaller systems at lower redshifts.

\subsection{High Redshifts ($z\ga 2$)}

As at $z\la 2$, at redshifts exceeding the peak of quasar activity
$(z\ga 2)$, the evolutions of the SFRD and BARD are, in fact, not
parallel (see Fig.~\ref{fig:madau}).  It must be noted that the
observational determinations of both quantities are much less certain
at these high redshift than at low redshifts. The presence of a
population of high--redshift, dust obscured quasars could, for
example, reconcile the SFRD and BARD curves in our simple model. There
is already evidence for such a population that could significantly
increase the inferred BARD (e.g. Fabian \& Iwasawa 1999); there is
also some evidence that, unlike the optical LF, the soft X--ray quasar
luminosity function stays flat out to redshifts $z\sim 4$ (Miyaji et
al. 2001).  We here simply take the current determinations at face
value, and examine physical reasons that would explain the apparent
discrepancy, if it holds up in future data.

\subsubsection{What Steepens the Evolution of the High-$z$ Quasar Emissivity?}

One possibility is that the fueling rate of quasars is suppressed by
intrinsic physical limits to the rate of accretion.  Models in which
the BHs shine with approximately their Eddington luminosity can
naturally explain the observed evolution of the QSO luminosity
function (Haiman \& Loeb 1997; Haehnelt, Natarajan \& Rees 1998;
Wyithe \& Loeb 2002), by associating the rise from $z=6$ to $z=2$ with
the increase in the nonlinear mass--scale in hierarchical cosmologies.
Ciotti \& Ostriker (2003) suggested that at the high characteristic
densities at $z\ga 3$, Bremsstrahlung opacity may effectively limit
the mass accretion rate onto a BH to a small fraction of the usual
Eddington value.  This idea is attractive because it provides a
physical reason for the suppression of the fueling rate, and because
the additional opacity may be relevant only at high redshifts,
allowing ``normal'' accretion at $z\la 3$. In fact, the data reviewed
in section \S~3.1.1 are consistent with $L_{\rm max} = 0.1 L_{\rm
Edd}$ (and a modest correction for beaming), but the effects of
correspondingly increasing the Eddington time by a factor of 10 are
only important at high redshifts, where it then becomes comparable to
the age of the universe.

In Figure~\ref{fig:corr}, we show (dot--dashed curve) the evolution of
the emissivity for a BH, $L \propto \dot M \propto M \propto
\exp(f_{\rm Edd} t/t_{\rm Edd})$ under the assumption that the hole
grows exponentially on a timescale which is $f_{\rm
Edd}^{-1}=1/0.07=14$ times the Eddington time $t_{\rm Edd}=\epsilon
\times 4.6\times10^8$yr. A multiplicative constant ($10^{-3.1}{\rm
M_\odot\, yr^{-1} Mpc^{-3}}$ for the curve shown in
Fig.~\ref{fig:corr}) can be used to represent the summed emissivity or
accretion rate density of all quasar BHs, all of which are assumed to
grow at the same rate from $z=\infty$.  As the figure reveals, the
suppression of the accretion rates in all BHs to $7\%$ of the
Eddington value would naturally result in the observed steep slope of
the quasar emissivity evolution between $3\la z\la 6$, while not
preventing star--formation to occur in a more extended spheroid region
around the black hole.  While attractive, this explanation suffers
from a drawback, namely the fact that if all BHs can accrete only at
10\% of the Eddington rate at $z>3.5$, then their e-folding time will
be $\sim 5\times 10^8$ years, making it apparently difficult to
explain how the large (few $\times 10^9$ ${\rm M_\odot}$) BHs in the
SDSS survey were built by $z=6$, when age of the universe is $8\times
10^8$ years, less than twice the $e$-folding timescale (Haiman \& Loeb
2001).  As we shall see below, this is less of a problem than might be
expected, since the growth of individual SMBHs at high redshift is
dominated by mergers, and not by accretion.

\subsubsection{The Growth of an Individual Black Hole due to Mergers vs. Accretion}

A different, but potentially important ingredient in determining the
relative evolution of the SFRD and the BARD at high redshifts is the
importance of mergers (see also Volonteri et al. 2003).  We next
demonstrate that at high redshifts, the build-up of the mass of an
individual BH is likely dominated by mergers between BHs.  Such
mergers may not have any effect on the total quasar emissivity (one
can imagine merging all BHs in pairs, resulting in new BHs twice as
massive as the original set, while preserving the accretion rate per
unit BH mass). However, if mergers are frequent, one can imagine that
this may help explain the high redshift discrepancy between the SFRD
and the BARD.  For example, one can imagine that a merger event at
high redshift delivers new gas and triggers star--formation (but may
not be able to increase the accretion rate onto BHs, per unit BH mass,
if this quantity is already Eddington limited).

The growth rate of BHs due to merging can be obtained from the
characteristic dark matter halo number density as follows:

\beq
\dot M_{\rm merg} = M_{\rm nl} \frac{\dot N_{\rm nl}}{N_{\rm nl}}.
\eeq

Here we define the nonlinear dark matter halo mass--scale $M_{\rm nl}$
at redshift $z$ as $\sigma(M_{\rm nl})g(z)=1$, where $\sigma(M)$ is
the r.m.s. mass fluctuation in spheres of mass $M$, and $g(z)$ is the
growth function at redshift $z$.  For simplicity, we define the space
density of halos as $N_{\rm nl}\equiv M_{\rm nl}\times dN/dM_{\rm
nl}$, where $dN/dM$ is the usual (comoving) halo mass function,
adopted here from Jenkins et al. (2001), evaluated at $M_{\rm
nl}$. Under this assumption,$N_{\rm nl}=N_{\rm nl}(M(t),t)$, and we
have the following time derivative:

\beq
\dot N_{\rm nl}=  M_{\rm nl}\frac{d^2 N}{dM_{\rm nl}dt} + \dot M_{\rm nl}\frac{dN}{dM_{\rm nl}}.
\eeq

The first term on the right hand side vanishes by definiton
($d^2N/dMdt\sim 0$ at the nonlinear mass-scale).  As a result, we find
that

\beq
\dot M_{\rm merg} = 
0.13\times 0.1\times 1.3\times 10^{-3}\times
M_{\rm nl} \frac{\dot N_{\rm nl}}{N_{\rm nl}} \sim 1.7\times10^{-5}\times\dot M_{\rm nl}.
\label{eq:mdotmerg}
\eeq

This last result reflects the fact that without any accretion, the
individual BH masses would grow only by coalescence of the BHs during
halo mergers, and therefore the typical BH mass would simply track the
nonlinear dark halo mass--scale.  In order to describe the growth of
BHs, rather than that of halos, we have assumed in
equation~(\ref{eq:mdotmerg}) that a fraction 0.13 of the total mass in
each halo is baryonic, the mass of stars is 10\% of the baryons, and
the mass of the central BH is $1.3\times 10^{-3}$ that of the stars.

We next find the growth of an individual BH due to accretion, using
the Madau \& Pozzetti (2000) star formation rate density, as follows:

\beq \dot M_{\rm acc} = 1.3\times 10^{-3}\times \frac{\dot \rho_{\rm
S}}{N_{\rm nl}}, \label{eq:mdotacc}
\eeq
where $\rho_{\rm S}$ is the comoving star--formation rate density as
given by equations~(\ref{eq:rhosuv}) and (\ref{eq:sfr}) above, and we
take $1.3\times 10^{-3}$ for the ratio of the BH mass to the spheroid
mass from Kormendy and Gebhardt (2001).

\begin{figure}
\begin{center}
\includegraphics[width=0.7\textwidth]{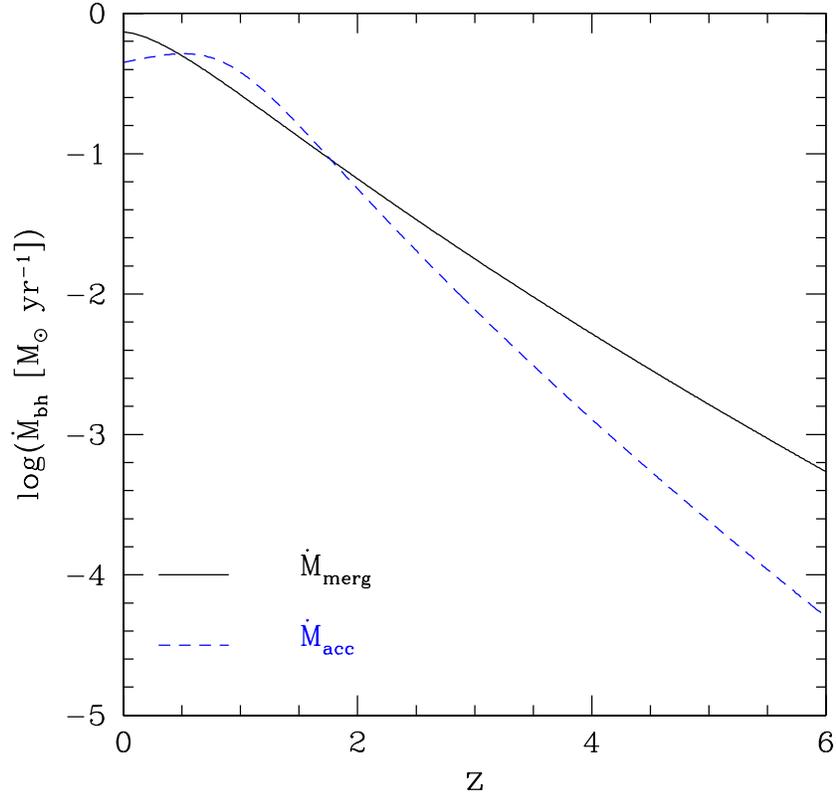}
\end{center}
\figcaption[fig3.eps]{The growth rate of an individual black hole with
the characteristic BH mass at each epoch, from accretion (dashed
curve) and mergers (solid curve), as a function of redshift.  Also
shown is the growth rate corresponding to accretion that is limited at
all times to a fraction 0.07 of the Eddington rate for the
characteristic BH mass at each epoch.\label{fig:growth}}
\end{figure}

Figure~\ref{fig:growth} shows the mass growth rates due to merging
(solid curve) and accretion (dashed curve).  From this figure, we
learn that at redshifts ($2<z<4$), the growth is dominated by mergers,
while at low redshift ($z<2$), the growth is proportional to the star
formation rate.\footnote{Note that at high redshifts ($z>4$), the mass
buildup by mergers actually exceeds $\sim 10\%$ of the Eddington rate,
in accordance with \S~3.2.1 above.}  According to the figure, at very
low redshifts ($z<0.5$), mergers are again important; however this
regime is unphysical because of the so--called ``over--merging''
problem in the Press--Schechter formalism: the nonlinear mass--scale
grows to that corresponding to clusters of galaxies; the galaxies,
however, may preserve their identities, and hence it is no longer
clear that the growth of BHs by coalescence in merging galaxies tracks
this mass-scale.

\section{Radiative Efficiency and Duty Cycle of AGNs}

In this section, we define and derive the radiative efficiency and the
duty cycle of AGNs; quantities that served as inputs in the previous
sections.  In the interest of clarity, let us first consider a
population of $\Ng$ identical galaxies over the Hubble time $\tH$,
each of which today (i.e. at $t=\tH$) hosts a spheroid of mass $\Ms$,
and a BH of mass $\Mbh$.  Let us further assume that during the entire
time elapsed from 0 to $\tH$, each BH had only two states: it was
either ``on'' or ``off''. We identify the ``on'' state as the active
quasar phase, and we define the duty cycle $\fq$ as the fraction of
the time each BH spends in the ``on'' state. At any given time, the
number of active quasars is then $\Nq=\fq\Ng$.  In the ``on'' state,
the BH grows by accretion at the rate $\dot\Mbh$, and shines at the
(bolometric) luminosity $\Lq$ with a radiative efficiency $\epsilon$,
defined as the fraction of the rest mass energy of the infalling gas
converted to radiation. The remaining fraction $(1-\epsilon)$ of the
rest mass then leads to the growth of the BH mass (Yu \& Tremaine
2002). A simple algebra shows that
\begin{equation}
\frac{\epsilon}{1-\epsilon} =
{\Etq\over \Mbht c^2}=
\frac{\fq \tH \Lq \Ng}{\Mbh \Ng c^2}=
\frac{\tH \Lq \Nq}{\Mbh \Ng c^2}.
\label{eq:toy_eps}
\end{equation}
Here $c$ is the speed of light; the numerator represents the total
light emitted by all BHs, and the denominator represents the total
mass in BHs today.  In the third equality, we have used
$\Nq=\fq\Ng$. Note that the last term involves only quantities that
are, in principle, directly observable, and that it is {\it
independent} of the duty cycle (Soltan 1982).  Equation
(\ref{eq:toy_eps}) describes the entire galaxy population, but a
similar equation applies to individual galaxies:
$\Lq\fq\tH=\epsq \Mbh c^2$.  This last expression does have a
dependence on the duty cycle, which can therefore be written as
\begin{equation}
\fq=\frac{\Nq}{\Ng}={\epsq\Mbh c^2\over \tH\Lq}
\label{eq:toy_duty}
\end{equation}

The simple toy model above demonstrates that (i) the radiative
efficiency can be obtained independently of the duty cycle, and (ii)
that the duty cycle can be obtained two different ways, based either
on the {\it number} or on the {\it characteristic BH mass} of quasars.
While the former method is conceptually more straightforward, as we
shall see below, the latter avoids the divergence in the number of
quasars and galaxies (due to the steep observed slope of the
luminosity functions at the faint end).

The step forward to a more realistic situation is to allow a
distribution of galaxy and BH masses, and corresponding quasar
luminosities, and to allow the BH masses and luminosities to change
with time.  In this case, equation~(\ref{eq:toy_eps}) can be
straightforwardly generalized to obtain the global average radiative
efficiency (Soltan 1982).  The total energy output of all quasars over
all times per unit volume is given by

\beq 
u_{\rm Q}= -A_{\rm bol} \int_0^{z_{\rm max}} dz \frac{dt}{dz}
\int_{L_{\rm min}}^{L_{\rm max}} dL\, L \Phiq(L,z)
\label{eq:epsq}
\eeq 
where $\Phiq(L,z)$ is the observed quasar luminosity function, which
we take from equation~(\ref{eq:phiq}) in the B--band, $A_{\rm bol}$ is
the bolometric correction, which, from the composite quasar spectrum
in Elvis et al. (1994), we find to be $A_{\rm bol}=L_{\rm
tot}/L_B=11.2$, and $dt/dz$ is obtained from the time--redshift
relation in our chosen $\Lambda$CDM cosmology. Note that the integral
in equation~(\ref{eq:epsq}) converges both from above and below in
luminosity, and in redshift, so that it is insensitive to the
integration limits $L_{\rm min}$, $L_{\rm max}$, and $z_{\rm max}$.
We find 
\beq 
\frac{u_{\rm Q}}{c^2}=1.9\times10^4\,{\rm M_\odot\,
Mpc^{-3}}
\label{eq:qsolight}
\eeq

The total remnant BH mass in the present universe is given by 

\beq 
\rho_{\rm BH}=\int_0^\infty dM M \Phi_{\rm BH}(M) = f_{\rm
BH}\int_0^\infty dM M \Phis(M),
\eeq 
where $\Phi_{\rm BH}$ and $\Phis$
are the present day BH and spheroid mass functions, and $f_{\rm BH}$
is the BH to spheroid mass ratio.  We here adopt the spheroid
luminosity function from equation~(\ref{eq:dphisphdl}), and convert it
to a spheroid mass function using the $M-L$ ratio for each of the 4
different types of galaxies in Salucci et al. (1999).  Assuming
further a constant BH--spheroid mass ratio $f_{\rm
BH}=1.3\times10^{-3}$ (Kormendy \& Gebhardt 2001), we find 
\beq
\rho_{\rm BH}=2.49\times 10^5\, {\rm M_\odot\, Mpc^{-3}}.
\label{eq:qsomass}
\eeq It is worth noting the contribution to this mass density from the
(E, S0, Sa/Sab, and Sbc/Scd) galaxies separately, which are,
respectively, 1.19, 0.63, 0.50, and 0.17 ${\rm M_\odot\, Mpc^{-3}}$.
In other words, the bulges of late type galaxies are not a negligible
contribution to the total spheroid mass density (and they dominate the
total mass density for low spheroid masses).  It is also worth noting
that the present--day total spheroid mass density we obtain by
integrating the spheroid formation rate in Figure~\ref{fig:corr}
(which was based on a corrected version of the evolution of the total
star formation rate in Madau \& Pozzetti 2000) is in good agreement
(to within 30\%) of the value we find from summing up the inferred
local spheroid densities from Salucci et al. (1999; see
eqn.~\ref{eq:dphisphdl}).

Finally, from the ratio of equations~(\ref{eq:qsolight}) and
(\ref{eq:qsomass}), we find $\epsq=0.071$.  Our result is in good
agreement with recent work by Yu \& Tremaine (2002), who find an
average efficiency of $\epsq=0.077$.  It is interesting to note that
this agreement holds separately for the QSO light and BH mass
density. Yu \& Tremaine use the recent 2dF quasar luminosity function
of Boyle et al. (2000), and a bolometric correction of 11.8, to find
$u_{\rm Q}/c^2=2.1\times10^4\,{\rm M_\odot\, Mpc^{-3}}$.  They combine
the recent $\Mbh-\sigma$ relation of Tremaine et al. (2002) with the
SDSS velocity function of early type galaxies in Bernardi et
al. (2002), and make corrections for the additional BH mass in spiral
galaxies and for the scatter in the $\Mbh-\sigma$ relation, yielding
$\rho_{\rm BH}=2.5\times 10^5\, {\rm M_\odot\, Mpc^{-3}}$.

The duty cycle defined in equation~(\ref{eq:toy_duty}) is less
trivially generalized for a population of evolving galaxies.
Nevertheless, if we assume that the duty cycle does not vary with time
(but we allow it to be a function of luminosity), then we can still
obtain the duty cycle explicitly by comparing the present--day space
density of quasars and galaxies.  An immediate complication is that
both space densities (see eqns. \ref{eq:dphisphdl} and \ref{eq:phiq})
diverge at the faint end. We therefore proceed by defining the average
duty cycle of all quasars above luminosity $\Lq(x,0)$,
\begin{equation}
\langle\fqN\rangle_x\equiv 
\frac{\int_{\Mbh(x)}^\infty dM \Phi_{\rm BH}}
     {\int_{\Lq(x,0)}^\infty dL \Phiq},
\label{eq:fdutyN}
\end{equation}
where $\Lq(x,0)$ is such that QSOs at redshift $z=0$ brighter than
$\Lq(x,0)$ emit a fraction $x$ of the total quasar light $\Lqt=
\int_{0}^{\infty} dL\, L \Phiq(L,0)$, and likewise, $\Mbh(x)$ is such
that all BHs more massive than $\Mbh(x)$ sum up to the same fraction
$x$ of the total BH mass at $z=0$, i.e. to $x\Mbht$.

We may similarly generalize the definition of the duty cycle based on
the characteristic BH mass (cf. the last term in
equation~[\ref{eq:toy_duty}]), by applying it to an individual
present--day BH, as follows:
\begin{equation}
\langle\fqM\rangle_x\equiv 
\frac{\epsilon c^2 \Mbh (x)}
     {\int_0^{\infty} \Lq (x,t)dt},
\label{eq:fdutyM}
\end{equation}
In principle, the time--integral in the denominator on the right-hand
side must be taken over the (typical) luminosity history of the
present--day black hole of mass $\Mbh$.  If mergers play an important
role in the growth of this BH, then the integral must be performed
separately, and summed over each branch along the ``merger tree''.  In
practice, we do not know this merger history, and instead we simply
define the time--dependent luminosity $\Lq(x,t)$ such that QSOs at
cosmic time $t$ brighter than $\Lq(x,t)$ emit a fraction $x$ of the
total quasar light at that epoch, $\Lqt=\int_{0}^{\infty} dL\, L
\Phiq(L,t)$.  This definition assumes only that a monotonic relation
is maintained between $\Mbh$ and $\Lq$ at all times, and it gives the
correct duty cycle in the limit that mergers do not dominate the mass
growth of individual BHs (note that the bulk of the total quasar light
is emitted in a narrow peak around redshift $z=2.2$; $\sim50\%$ of the
time--integral in equation~(\ref{eq:fdutyM}) is contributed between
$1.6<z<2.8$).  As we showed in the previous section (see
Figure~\ref{fig:growth}), mergers are likely important at $z>3$ but do
not dominate the growth of individual BHs at lower redshifts.

The duty cycles obtained from both methods are listed in Table 1.  We
find that $\langle\fqN\rangle_{0.1}\simeq 0.008$ and
$\langle\fqN\rangle_{0.9}\simeq 0.05$. Most importantly, our two
definitions above do not guarantee that equations~(\ref{eq:fdutyN})
and ~(\ref{eq:fdutyM}) give the same values.  Here we find that the
two methods agree well on the high mass end, while
$\langle\fqM\rangle$ is systematically lower by a factor of $\sim$two
towards the low--mass end.  One interpretation of this finding is that
the most massive BHs gain nearly all their mass by accretion, while
mergers contribute a comparatively larger fraction of the mass of
lower mass BHs.

Our results for the duty cycle being at the percent level is in good
agreement with theoretical expectations (Ciotti \& Ostriker 1997,
2001, Yu \& Tremaine 2002), it is also similar to the values derived
in Haehnelt, Natarajan, \& Rees (1998), who obtained a {\it QSO
lifetime} of $t_{\rm Q}\simeq 10^7$ yr at a Hubble epoch of $\simeq
10^{9-10}$ yr or, in terms of the duty cycle, $\fq=t_{\rm Q}/t_{\rm
Hubble}\simeq 10^{-2}-10^{-3}$.  Similar quasar lifetimes can be
independently derived from the spatial clustering of quasars (Haiman
\& Hui 2001; Martini \& Weinberg 2001).

Before we conclude this section, we stress a well known puzzle
presented by the difference in the QSO luminosity functions in the
optical and X--ray bands, and its consequences for our analysis. We
repeated the derivation of the accretion efficiency from the X--ray
luminosity function (Miyaji et al. 2001), and we find $\epsq\sim
0.045$, about two times lower than we obtain from the B--band
(applying in both cases a bolometric correction from Elvis et
al. 1994: $L_{\rm tot}/L_X=38.1$, and $L_{\rm tot}/L_B=11.2$). If all
QSOs emitted intrinsically with the Elvis spectrum, the above two
numbers should agree.  However, they differ by a factor of two.  What
does this mean?  The simplest resolution is to assume that QSOs emit a
universal spectrum, but the mean value of their flux ratio $L_X/L_B$
is a factor of $71/45=1.6$ times smaller than for the Elvis et
al. sample (which consists of 47 unobscured QSOs). This says
$L_X/L_B\approx 0.18$ .  However, under the assumption of a universal
spectrum, the total number of quasars \beq N(>L_{\rm min})=
4\pi\int_0^\infty dz \frac{dV}{dz d\Omega} \int_{L_{\rm min}}^\infty
dL\, \frac{d\Phi}{dL}(L,z) \eeq should then be equal in the optical
and in the X--rays, if one uses the appropriate lower limits, $L_{\rm
min,X}$=$0.18 L_{\rm min,B}$.  We find that this leads instead to
$N_X=2 N_B$.  In other words, under the assumption of a universal
spectrum, the flux ratio $L_X/L_B$ can be derived two ways: either
from the total number, or from the total light, of quasars. With the
published LFs, these methods give $L_X/L_B\approx 0.5$ and
$L_X/L_B\approx 0.2$, respectively.  This proves that quasars as a
population cannot have a universal spectrum, a result that one could
have obtained directly by comparing the X-ray and optical LFs, which
have different shapes and redshift evolutions.  A possible resolution
is that the duty-cycle in the X-rays is $\sim 3$ times longer than in
the optical.  This would be supported by the recent results of Barger
et al. (2001), who found that a large fraction of optical galaxies are
{\it Chandra} AGN sources, implying a long X--ray activity cycle,
$\sim$0.5 Gyr.

\section{Conclusions}

In this paper, we discussed a simple, empirically based model for the
growth of supermassive black holes (BHs) at the center of spheroidal
stellar systems. Motivated by accumulating evidence for the strong
link between the formation of spheroids and BHs, we hypothesized the
{\it simplest possible form of this connection}, namely that star
formation in spheroids and BH fueling are proportional to one another,
at all cosmic epoch and in all spheroids, regardless of their size.

The main conclusions that arise from this hypothesis (augmented with a
few other reasonable assumptions) are as follows.  This simple model
accounts for the puzzling steep decline of the characteristic
luminosity of quasars from redshift $z\approx 2$ to $z=0$: the
residual star formation in spheroidal systems is today limited to a
small number of bulges, characterized by stellar velocity dispersions
a factor of $2-3$ smaller those of the elliptical galaxies hosting
QSOs at $z \ga 2$.  We explored a very simple consequence of our
hypothesis: the redshift evolution of the QSO emissivity and of the
star formation history in spheroids should be roughly parallel to each
other. We find this result to be broadly consistent with the evolution
of both the global star formation rate, and of the evolution of the
QSO emissivity, both of which exhibit a peak at redshift $z\sim 2$.
However, a closer look reveals possibly interesting discrepancies at
both low and high redshifts.  

At low redshifts, the spheroid formation rate, obtained by making
simple corrections to the total star formation rate, appears to
decline by a factor that is $\sim 3$ times larger than the decline in
QSO emissivity.  A possible solution we note to resolve this
discrepancy is fueling of quasar BHs at low redshifts by the mass lost
in winds from a passively evolving stellar spheroid population, formed
at earlier epochs.  A tentative discrepancy also exists at high
redshifts ($z\ga 2$, beyond the peak of QSO activity), where the
evolution of the star formation rate appears significantly flatter
than that of quasar emissivity.  While a population of hitherto
undetected, obscured AGN at high redshift (with the obscured fraction
increasing towards high $z$) could resolve this discrepancy, we
offered an alternative, physical explanation: quasar fueling rates at
high redshift are limited to a fraction $\sim 10\%$ of the Eddington
accretion rate.  This limit depends linearly on the characteristic BH
mass, and would therefore imprint a steep evolution of the quasar
luminosity function as the characteristic mass--scale builds up
exponentially.  We also note that the masses of individual black holes
at high redshift are not limited by accretion (at the Eddington or
some modified Eddington rate), since we find that mergers dominate
over accretion in determining the growth of objects at epochs $z \gsim
2$.

Given our demographic assumptions, we compute the average duty cycle -
the fraction of time SMBHs spend in the on state - as $(3-6)\times
10^{-3}$, depending on BH mass, and we also find the mean bolometric
radiative efficiency, $\epsilon=0.071$, when averaged for the entire
SMBH population.

The considerations on this paper are tentative, but empirically based.
It should soon be possible to considerably tighten constraints on the
simple picture of coeval formation of BHs and spheroids: the
redshift--evolution of the starformation rate in spheroid systems
should be derivable by a more detailed analysis of the SDSS galaxy
sample out to at least $z\sim 0.5$.

\acknowledgments

We thank Martin Haehnelt, Martin Rees, Alvio Renzini, Scott Tremaine
and Qingjuan Yu for useful discussions.  L.C. was supported by MURST,
contract CoFin2000, and by grant ASI I/R/105/00.

\clearpage
\begin{deluxetable}{crrrrrrrr}
\footnotesize
\tablecaption{\label{tbl-1}}
\tablewidth{0pt}
\tablehead{
           \colhead{$x$} & 
           \colhead{$\Mbh(x)$} & 
           \colhead{$\Nbh(x)$} & 
           \colhead{$\Lq(x,0)/L_{\rm S*}$} &
           \colhead{$\Nqp (x,0)$} &
           \colhead{$\langle\fqN\rangle_x$} &
           \colhead{$\langle\fqM\rangle_x$} &
          }
\startdata

0.1 & 4.33e8 & 3.92e-5 & 1.51e0  & 3.02e-7 & 0.0077 & 0.00349\\ 
0.2 & 2.67e8 & 1.14e-4 & 0.80e0  & 1.05e-6 & 0.0092 & 0.00404\\
0.3 & 1.82e8 & 2.29e-4 & 0.49e0  & 2.33e-6 & 0.0102 & 0.00452\\
0.4 & 1.27e8 & 3.94e-4 & 0.30e0  & 4.40e-6 & 0.0112 & 0.00517\\
0.5 & 8.91e7 & 6.28e-4 & 0.18e0  & 7.85e-6 & 0.0125 & 0.00616\\
0.6 & 6.03e7 & 9.69e-4 & 0.93e-1 & 1.40e-6 & 0.0145 & 0.00784\\
0.7 & 3.76e7 & 1.50e-3 & 0.42e-1 & 2.67e-6 & 0.0179 & 0.0109\\
0.8 & 2.00e7 & 2.40e-3 & 0.14e-1 & 6.03e-5 & 0.0251 & 0.0180\\
0.9 & 7.05e7 & 4.48e-3 & 0.20e-2 & 2.19e-4 & 0.0488 & 0.0433\\

\enddata \tablecomments{The columns show respectively, the BH mass,
the BH space density, quasar luminosity, local QSO space density, and
duty cycles (last two columns, computed using two different methods),
for galaxies that contain a fixed fraction $x$ (from column 1) of
quasar light and both spheroid and BH mass.}
\end{deluxetable}

\clearpage

\clearpage

\end{document}